\input harvmac
%

\def\hat{\widehat}

%
\let\includefigures=\iftrue
%
%
%
\newfam\black
\input rotate
\input epsf
\noblackbox

\includefigures
\message{If you do not have epsf.tex (to include figures),}
\message{change the option at the top of the tex file.}
\def\figin{\epsfcheck\figin}\def\figins{\epsfcheck\figins}
\def\epsfcheck{\ifx\epsfbox\UnDeFiNeD
\message{(NO epsf.tex, FIGURES WILL BE IGNORED)}
\gdef\figin##1{\vskip2in}\gdef\figins##1{\hskip.5in}
\else\message{(FIGURES WILL BE INCLUDED)}%
\gdef\figin##1{##1}\gdef\figins##1{##1}\fi}
\def\DefWarn#1{}

\def\figinsert{\goodbreak\midinsert}
\def\ifig#1#2#3{\DefWarn#1\xdef#1{fig.~\the\figno}
\writedef{#1\leftbracket fig.\noexpand~\the\figno}%
\figinsert\figin{\centerline{#3}}\medskip\centerline{\vbox{\baselineskip12pt
\advance\hsize by -1truein\noindent\footnotefont{\bf
Fig.~\the\figno:} #2}}
\bigskip\endinsert\global\advance\figno by1}
\else
\def\ifig#1#2#3{\xdef#1{fig.~\the\figno}
\writedef{#1\leftbracket fig.\noexpand~\the\figno}%
\global\advance\figno by1} \fi
\def\yboxit#1#2{\vbox{\hrule height #1 \hbox{\vrule width #1
\vbox{#2}\vrule width #1 }\hrule height #1 }}
\def\fillbox#1{\hbox to #1{\vbox to #1{\vfil}\hfil}}
\def\ybox{{\lower 1.3pt \yboxit{0.4pt}{\fillbox{8pt}}\hskip-0.2pt}}

\def\rightarrowbox#1#2{
  \setbox1=\hbox{\kern#1{${ #2}$}\kern#1}
  \,\vbox{\offinterlineskip\hbox to\wd1{\hfil\copy1\hfil}
    \kern 3pt\hbox to\wd1{\rightarrowfill}}}

\def\vev#1{\langle{#1}\rangle}

\def\CA{{\cal A}}

\def\CJ{{\cal J}}

\def\CP{{\cal P}}

\def\tilde{\widetilde}

\def\II{\relax{I\kern-.10em I}}

\def\bar{\overline}

\def\IZ{\relax\ifmmode\mathchoice
{\hbox{\cmss Z\kern-.4em Z}}{\hbox{\cmss Z\kern-.4em Z}}
{\lower.9pt\hbox{\cmsss Z\kern-.4em Z}} {\lower1.2pt\hbox{\cmsss
Z\kern-.4em Z}}\else{\cmss Z\kern-.4em Z}\fi}
\def\IB{\relax{\rm I\kern-.18em B}}
\def\IC{{\relax\hbox{$\inbar\kern-.3em{\rm C}$}}}
\def\ID{\relax{\rm I\kern-.18em D}}
\def\IE{\relax{\rm I\kern-.18em E}}
\def\IF{\relax{\rm I\kern-.18em F}}
\def\IG{\relax\hbox{$\inbar\kern-.3em{\rm G}$}}
\def\IGa{\relax\hbox{${\rm I}\kern-.18em\Gamma$}}
\def\IH{\relax{\rm I\kern-.18em H}}
\def\II{\relax{\rm I\kern-.18em I}}
\def\IK{\relax{\rm I\kern-.18em K}}
\def\IN{\relax{\rm I\kern-.18em N}}
\def\IP{\relax{\rm I\kern-.18em P}}

%
\def\inbar{\,\vrule height1.5ex width.4pt depth0pt}

\font\cmss=cmss10 \font\cmsss=cmss10 at 7pt
\def\IR{\relax{\rm I\kern-.18em R}}

\def\lp10{l_P^{10}}
\def\lp11{l_P^{11}}
\def\R11{R_{11}}

%
%

%
%
%

\lref\LuoRX{
  M.~x.~Luo and C.~k.~Wen,
  ``Recursion relations for tree amplitudes in super gauge theories,''
  hep-th/0501121.
}

\lref\LuoMY{
  M.~x.~Luo and C.~k.~Wen,
  ``Compact formulas for all tree amplitudes of six partons,''
  hep-th/0502009.
}

\lref\BerendsZP{ F.~A.~Berends, W.~T.~Giele and H.~Kuijf, ``On
Relations Between Multi - Gluon And Multigraviton Scattering,''
Phys.\ Lett.\ B {\bf 211}, 91 (1988).
}

\lref\Bernbox{Z.~Bern, N.~E.~J.~Bjerrum-Bohr and D.~C.~Dunbar,
''Inherited Twistor-Space Structure of Gravity Loop Amplitudes,''
hep-th/0501137.}

\lref\BernBB{ Z.~Bern, N.~E.~J.~Bjerrum-Bohr and D.~C.~Dunbar,
``Inherited Twistor-Space Structure of Gravity Loop Amplitudes,''
hep-th/0501137.
}

\lref\BernHS{ Z.~Bern, L.~J.~Dixon and D.~A.~Kosower, ``On-shell
recurrence relations for one-loop QCD amplitudes,''
hep-th/0501240.
}

\lref\BernBT{ Z.~Bern, L.~J.~Dixon and D.~A.~Kosower, ``All
next-to-maximally helicity-violating one-loop gluon amplitudes in
N = 4 super-Yang-Mills theory,'' hep-th/0412210.
}

\lref\BernCG{ Z.~Bern, L.~J.~Dixon, D.~C.~Dunbar and
D.~A.~Kosower, ``Fusing gauge theory tree amplitudes into loop
amplitudes,'' Nucl.\ Phys.\ B {\bf 435}, 59 (1995) hep-ph/9409265.
}

\lref\BernSV{ Z.~Bern, L.~J.~Dixon, M.~Perelstein and
J.~S.~Rozowsky, ``Multi-leg one-loop gravity amplitudes from gauge
theory,'' Nucl.\ Phys.\ B {\bf 546}, 423 (1999) hep-th/9811140.
}

\lref\BernUG{ Z.~Bern, L.~J.~Dixon, D.~C.~Dunbar, M.~Perelstein
and J.~S.~Rozowsky, ``On the relationship between Yang-Mills
theory and gravity and its
Nucl.\ Phys.\ B {\bf 530}, 401 (1998) hep-th/9802162.
}

\lref\BernKY{ Z.~Bern, V.~Del Duca, L.~J.~Dixon and D.~A.~Kosower,
``All non-maximally-helicity-violating one-loop seven-gluon
amplitudes in N = 4 super-Yang-Mills theory,'' hep-th/0410224.
}

\lref\BernZX{ Z.~Bern, L.~J.~Dixon, D.~C.~Dunbar and
D.~A.~Kosower, ``One loop n point gauge theory amplitudes,
unitarity and collinear limits,'' Nucl.\ Phys.\ B {\bf 425}, 217
(1994) hep-ph/9403226.
}

\lref\BrittoAP{ R.~Britto, F.~Cachazo and B.~Feng, ``New recursion
relations for tree amplitudes of gluons,'' hep-th/0412308.
}

\lref\BrittoFQ{ R.~Britto, F.~Cachazo, B.~Feng and E.~Witten,
``Direct proof of tree-level recursion relation in Yang-Mills
theory,'' hep-th/0501052.
}

\lref\BrittoNC{ R.~Britto, F.~Cachazo and B.~Feng, ``Generalized
unitarity and one-loop amplitudes in N = 4 super-Yang-Mills,''
hep-th/0412103.
}

\lref\CachazoKJ{ F.~Cachazo, P.~Svr\v{c}ek and E.~Witten, ``MHV
vertices and tree amplitudes in gauge theory,'' JHEP {\bf 0409},
006 (2004) hep-th/0403047.
}

\lref\CremmerDS{ E.~Cremmer and B.~Julia, ``The N=8 Supergravity
Theory. 1. The Lagrangian,'' Phys.\ Lett.\ B {\bf 80}, 48 (1978).
}

\lref\CremmerKM{ E.~Cremmer, B.~Julia and J.~Scherk,
``Supergravity Theory In 11 Dimensions,'' Phys.\ Lett.\ B {\bf
76}, 409 (1978).
}

\lref\DeWittUC{ B.~S.~DeWitt, ``Quantum Theory Of Gravity. Iii.
Applications Of The Covariant Theory,'' Phys.\ Rev.\  {\bf 162},
1239 (1967).
}

\lref\HoweUI{ P.~S.~Howe and K.~S.~Stelle, ``Supersymmetry
counterterms revisited,'' Phys.\ Lett.\ B {\bf 554}, 190 (2003)
hep-th/0211279.
}

\lref\KawaiXQ{ H.~Kawai, D.~C.~Lewellen and S.~H.~H.~Tye, ``A
Relation Between Tree Amplitudes Of Closed And Open Strings,''
Nucl.\ Phys.\ B {\bf 269}, 1 (1986).
}

\lref\NairGR{V.~P.~Nair,''A Note on MHV Amplitudes for
Gravitons,'' hep-th/0501143.}

\lref\RoibanIX{ R.~Roiban, M.~Spradlin and A.~Volovich,
``Dissolving N = 4 loop amplitudes into QCD tree amplitudes,''
hep-th/0412265.
}

\lref\WittenNN{ E.~Witten, ``Perturbative gauge theory as a string
theory in twistor space,'' Commun.\ Math.\ Phys.\  {\bf 252}, 189
(2004) hep-th/0312171.
}

\lref\BedfordYY{ J.~Bedford, A.~Brandhuber, B.~Spence and
G.~Travaglini, ``A recursion relation for gravity amplitudes,''
hep-th/0502146.
}

\newbox\tmpbox\setbox\tmpbox\hbox{\abstractfont
}
 \Title{\vbox{\baselineskip12pt\hbox to\wd\tmpbox{\hss
 hep-th/0502160} }}
 {\vbox{\centerline{Tree Level Recursion Relations }
 \smallskip
 \centerline{In General Relativity}
 }}
\smallskip
\centerline{Freddy Cachazo and Peter Svr\v{c}ek\footnote{$^1$}{On
Leave from Princeton University}}
\smallskip
\bigskip
\centerline{{\it  Institute for Advanced Study, Princeton NJ 08540
USA }}
\bigskip
\vskip 1cm

\noindent

Recently, tree-level recursion relations for scattering amplitudes
of gluons in Yang-Mills theory have been derived. In this note we
propose a generalization of the recursion relations to tree-level
scattering amplitudes of gravitons. We use the relations to derive
new simple formulae for all amplitudes up to six gravitons. In
particular, we present an explicit formula for the six graviton
non-MHV amplitude. We prove the relations for MHV and next-to-MHV
$n$-graviton amplitudes and for all eight-graviton amplitudes.

\Date{February 2005}


\newsec{Introduction}

Lately, there has been a lot renewed progress in understanding the
tree-level and one-loop gluon scattering amplitudes in Yang-Mills
theory. Among other things, a new set of recursion relations for
computing tree-level amplitudes of gluons have been recently
introduced in \BrittoAP.\ A proof of the recursion relations was
given in \BrittoFQ.\ A straightforward application of these
recursion relations gives new and simple forms for many amplitudes
of gluons. Many of these have been obtained recently using
somewhat related methods \refs{\BernKY, \BernBT, \RoibanIX}.
Application of recursion relations to amplitudes with fermions
also lead to very compact formulas \refs{\LuoRX, \LuoMY}.

It has been known that tree level graviton amplitudes have
remarkable simplicity that cannot be expected from textbook
recipes for computing them. The tree level $n$ graviton amplitudes
vanish if more than $n-2$ gravitons have the same helicity. The
maximally helicity violating (MHV) amplitudes are thus, as in
Yang-Mills case, those with $n-2$ gravitons of one helicity and
two of the opposite helicity. These have been computed by Berends,
Giele, and Kuijf (BGK) \BerendsZP\ from the Kawai, Lewellen and
Tye (KLT) relations \KawaiXQ.\ The four particle case was first
computed by DeWitt \DeWittUC.\

This raises the question whether there are analogous recursion
relations for amplitudes of gravitons which would explain some of
the simplicity of the tree-level graviton amplitudes. The
possibility of such recursion relations has been recently raised
in \BernBB.\

In this note, we propose tree-level recursion relations for
amplitudes of gravitons. The recursion relations can be
schematically written as follows \eqn\grsh{A_n=\sum_{{\cal
I},h}A_{\cal I}^h{1\over P_{\cal I}^2}A^{-h}_{\cal J}.} Here $A$
denotes a tree level graviton amplitude. In writing a recursion
relation for $n$ graviton amplitude $A_n,$ one marks two gravitons
and sums over products of subamplitudes with external gravitons
partitioned into sets ${\cal I}\cup {\cal J}=(1,2,\dots,n),$ among
the two subamplitudes so that $i\in\CI$ and $j\in\CJ$. $P_{\cal
I}$ is the sum of the momenta of gravitons in the set ${\cal I}$
and $h$ is the helicity of the internal graviton. The momenta of
the internal and the marked gravitons are shifted so that they are
on-shell.

We use the recursion relations to derive new compact formulas for
all amplitudes up to six gravitons. In particular, we give the
first published result for the six graviton non-MHV amplitude
$A(1^-,2^-,3^-,4^+,5^+,6^+).$

We attempt to prove the recursion relations along the lines of
\BrittoFQ.\ The first part of the proof that rests on basic facts
about tree-level diagrams, such as the fact that their
singularities come only from the poles of the internal propagators
can be easily adapted to the gravity case. To have a complete
proof of the recursion relations, it is necessary to prove that
certain auxiliary function $A(z)$ constructed from the scattering
amplitude vanishes as $z\rightarrow\infty.$

We are able to prove this fact from the KLT relations for all
amplitudes up to eight gravitons. For amplitudes with nine or more
gravitons, the KLT relations suggest that the function $A(z)$ does
not vanish at infinity unless there is an unexpected cancellation
between different terms in the KLT relations.

While we are not able to prove that $A(z)$ vanishes at infinity
for a general $n$ graviton scattering amplitude, we show that
$A(z)$ does vanish at infinity for MHV amplitudes with arbitrary
number of gravitons from the BGK formula. Hence, the recursion
relations are valid for all MHV amplitudes contrary to the
expectation from KLT relations.

Finally, we introduce an auxiliary set of recursion relations for
NMHV amplitudes which are easier to prove but give more
complicated results for the amplitudes. This auxiliary recursion
relation is then used to prove the vanishing of $A(z)$ for any
NMHV amplitudes.

This raises the hope, that the recursion relations hold for other
scattering amplitudes of gravitons as well.


\newsec{Recursion Relations}

Just like gauge theory scattering amplitudes, the graviton
scattering amplitudes are efficiently written in terms of
spinor-helicity formalism. In a nutshell, a on-shell momentum of a
massless particle can be written as a bispinor, $p_{a\dot
a}=\lambda_a\tilde\lambda_{\dot a}.$ Here, $\lambda$ and
$\tilde\lambda$ are commuting spinors with invariant products
$\vev{\lambda,\lambda'}=\epsilon_{ab}\lambda^a\lambda'{}^b$ and
$[\tilde\lambda,\tilde\lambda']=\epsilon_{\dot a \dot
b}\tilde\lambda^{\dot a}\tilde\lambda'{}^{\dot b}.$ Notice that
$2p\cdot p'=\vev{\lambda,\lambda'}[\tilde\lambda,\tilde\lambda'].$
We will also find it convenient to introduce the symbol $\langle
\lambda|p|\tilde\lambda']=\lambda^a p_{a\dot
a}\tilde\lambda'{}^{\dot a}.$ The polarization tensors of the
gravitons can be expressed in terms of gluon polarization vectors
\eqn\grpol{\epsilon^+_{a\dot a,b\dot b}=\epsilon^+_{a\dot
a}\epsilon^+_{b\dot b}\qquad \epsilon^-_{a\dot a,b\dot
b}=\epsilon^-_{a\dot a}\epsilon^-_{b\dot b}.} The polarization
vectors of positive and negative helicity gluons  are respectively
\eqn\glpol{\epsilon^-_{a\dot a}={\lambda_a\tilde\mu_{\dot a}\over
[\tilde\lambda,\tilde\mu]}\qquad \epsilon^+_{a\dot
a}={\mu_a\tilde\lambda_{\dot a}\over \vev{\mu,\lambda}},} where
$\mu$ and $\tilde\mu$ are fixed reference spinors.

Consider a tree level graviton scattering amplitude
$A(1,2,\dots,n).$ The amplitude is invariant under any
permutations of the gravitons because there is no color ordering.

To write down the recursion relations, we single out two
gravitons. Without loss of generality, we call these gravitons $i$
and $j.$  Define the shifted momenta $p_i(z)$ and $p_j(z),$ where
$z$ is a complex parameter, to be
\eqn\shiftp{p_i(z)=\lambda_i(\tilde\lambda_i+z\tilde\lambda_j)\qquad
p_j(z)=(\lambda_j-z\lambda_i)\tilde\lambda_j.} Note that $p_i(z)$
and $p_j(z)$ are on-shell for all $z$ and that
$p_i(z)+p_j(z)=p_i+p_j.$ Hence, the following function
\eqn\zfunction{A(z)=A(p_1,\dots,p_i(z),\dots,p_j(z),\dots,n)} is a
physical on-shell scattering amplitude for all values of $z.$

Consider the partitions of the gravitons $(1,2,\dots, i,\dots,
j,\dots,n)={\cal I}\cup {\cal J}$ into two groups such that $i\in
{\cal I}$ and $j\in {\cal J}.$ Then the recursion relation for a
tree-level graviton amplitude is \eqn\grtree{\eqalign{A(z)=
\sum_{{\cal I},{\cal J}}\sum_h A_L({\cal I},-P_{\cal I}^h(z_{\cal
I}),z_{\cal I}){1\over P_{\cal I}^2(z)} A_R({\cal J},P_{\cal
I}^{-h}(z_{\cal I}),z_{\cal I}),}} where
\eqn\defr{\eqalign{P_{\cal I}(z)&=\sum_{k\in {\cal I}, k\neq i}
p_k+p_i(z) \cr z_{\cal I}&={P_{\cal I}^2\over \langle i|P_{\cal
I}|j]}.}}  The sum in \grtree\ is over the partitions of gravitons
and over the helicities of the intermediate gravitons. The
physical amplitude is obtained by taking $z$ in equation \grtree\
to be zero \eqn\aphys{A(1,2,\dots, n)=A(0).}

We will give evidence below that the recursion relation is valid
for gravitons $i$ and $j$ of helicity $(+,+),(-,-)$ and $(-,+)$
respectively.

\ifig\recrel{This is a schematic representation of the recursion
relations \grtree.\ The thick lines represent the reference
gravitons. The sum here is over all partitions of the gravitons
into two groups with at least two gravitons on each subamplitude
and over the two choices of the helicity of the internal
graviton.} {\centerline {\epsfysize=1.in
\epsfbox{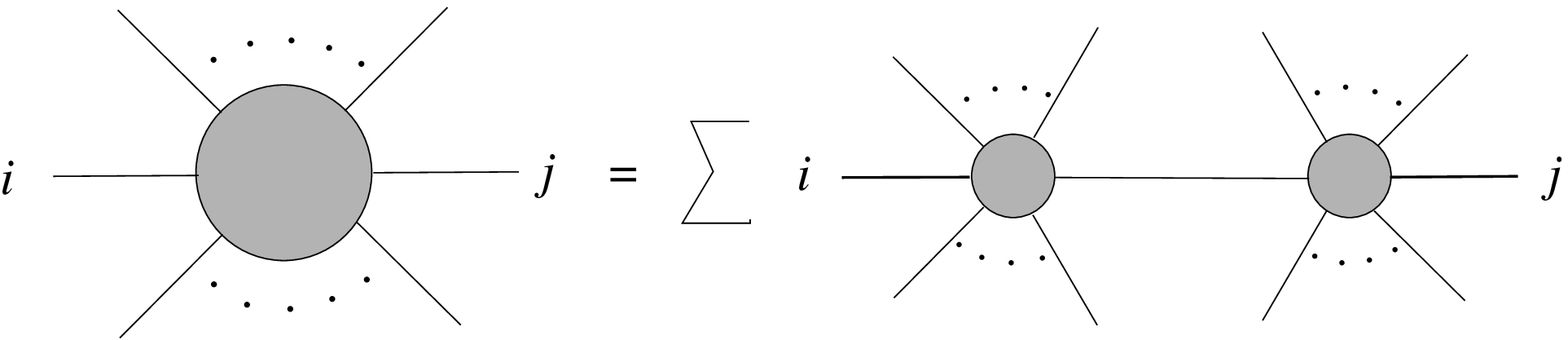}}}


\newsec{Explicit Examples}

In this section, we compute all tree-level amplitudes up to six
gravitons to illustrate the use of the recursion relations
\grtree.\

Consider first the four-graviton MHV amplitude
$A(1^-,2^-,3^+,4^+).$ The amplitude is invariant under arbitrary
permutations of external gravitons so the order of gravitons does
not matter. Hence, this is the only independent four graviton
amplitude. In contrast, in gauge theory, there are two independent
amplitudes $A_{YM}(1^-,2^-,3^+,4^+)$ and $A_{YM}(1^-,3^+,2^-,4^+)$
because the Yang-Mills scattering amplitudes are color ordered.

\ifig\four{Two configurations contributing to the four graviton
amplitude $A(1^-,2^-,3^+,4^+).$ Notice that the diagrams are
related by the interchange $2\leftrightarrow 3.$ }
{\epsfysize=1.2in\epsfbox{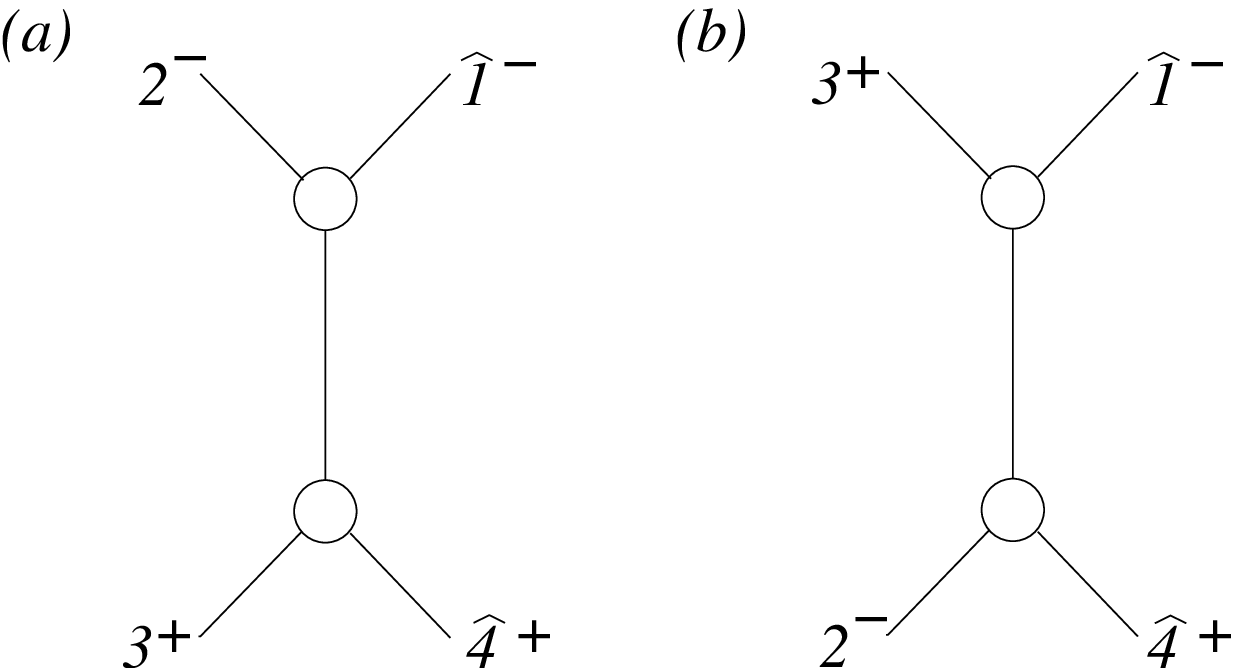}}

We single out gravitons $1^-$ and $4+.$   Then, there are two
possible configurations contributing to the recursion relations
\grtree,\ see \four.\ We refer to the configuration  from fig.
2(a) as $(2,\hat 1|\hat 4, 3) $ and from fig. 2(b) as $(3,\hat
1|\hat 4, 3).$ To evaluate the diagrams we use the known form of
three graviton scattering amplitudes
\eqn\threegr{A(1^-,2^-,3^+)={\vev{12}^6\over\vev{23}^2\vev{31}^2},
\qquad A(1^+,2^+,3^-)={[12]^6\over [23]^2[31]^2}.}

The sum of the two contributions  from fig. 2 is
\eqn\grfour{A(1^-,2^-,3^+,4^+)={\vev{12}^5[34]^2\over
[12]\vev{23}^2\vev{14}^2}+{\vev{12}^8[24]^2\over
\vev{13}^3[13]\vev{23}^2\vev{14}^2}.} A short calculation shows
that \grfour\ equals to the known result \BerendsZP\ obtained from
KLT relations \eqn\grgiele{A(1^-,2^-,3^+,4^+)={\vev{12}^8[12]\over
\vev{12}\vev{13}\vev{14}\vev{23}\vev{24}\vev{34}^2}.}

We picked the reference gravitons to have opposite helicity
because this leads to most compact expressions for graviton
scattering amplitudes. We could have chosen reference gravitons of
the same helicity, ie. $1^-$ and $2^-.$ This leads to a longer
expression  because there are more diagrams contributing to the
scattering amplitude. In the rest of the paper, we will always
choose reference gravitons of opposite helicity. The actual choice
of reference gravitons does not matter, because the amplitude is
invariant under permutations that preserve the sets of positive
and negative helicity gravitons. All choices lead to the same
answer up to relabelling of the gravitons.

The next amplitude to consider is the five graviton MHV amplitude
$A(1^-,2^-,3^+,4^+,5^+).$ Just as in the four graviton example,
this is the only independent five graviton amplitude. All other
five graviton amplitudes are related to it by permutation and/or
conjugation symmetry.

The amplitude has contribution from three diagrams $(1,4,\hat
2|\hat 3,5), (1,5,\hat 2|\hat 3, 4), (4,5,\hat 2|\hat 3, 1).$
These contributions give the following three terms
\eqn\fivegr{\eqalign{A(1^-,2^-,3^+,4^+,5^+)
={\vev{12}^7\over\vev{14}\vev{15}\vev{23}^2\vev{45}}\left({[14][35]
\over\vev{24}\vev{35}}-{[15][34]\over\vev{25}\vev{34}}-
{\vev{12}[13][45]\over\vev{13}\vev{24}\vev{25}}\right).}} This
expression agrees with the BGK result \BerendsZP\
\eqn\bgkfive{A(1^-,2^-,3^+,4^+,5^+)=\vev{12}^7{([12]\vev{23}[34]\vev{41}
-\vev{12}[23]\vev{34}[41])\over\vev{13}\vev{14}\vev{15}\vev{23}\vev{24}
\vev{25}\vev{34}\vev{35}\vev{45}}.}

At six gravitons, there are two independent scattering amplitudes,
the MHV amplitude $A(1^-,2^-,3^+,4^+,5^+,6^+)$ and the first
non-MHV amplitude $A(1^-,2^-,3^-,4^+,5^+,6^+).$

The MHV amplitude $A(1^-,2^-,3^+,4^+,5^+,6^+)$  has contribution
from four configurations,  $(4,\hat 3|\hat 2,1,5,6), (5,\hat
3|\hat 2,1,4,6),(6,\hat 3|\hat 2,1,4,5)$ and $(1,\hat 3|\hat
2,4,5,6).$ Notice that the first three diagrams are related by
interchange of $4,5,6$ gravitons, so there are only two diagrams
to compute.

The first configuration $(4,\hat 3|\hat 2,1,5,6)$ evaluates to
\eqn\secondfive{D_1=\vev{12}^7[34]{\langle2|3+4|5]\langle4|2+3|1]\vev{51}
-\vev{12}p_{234}^2\vev{45}[51]\over
\vev{14}\vev{15}\vev{16}\vev{23}^2\vev{25}\vev{26}\vev{34}\vev{45}\vev{46}\vev{56}}.}

The last configuration  $(1,\hat 3|\hat 2,4,5,6)$ gives
\eqn\firstfive{D_2=\vev{12}^8[13]{\vev{14}[45]\vev{52}p_{123}^2
-\vev{45}\langle2|1+3|4]\langle1|2+3|5]\over
\vev{13}\vev{14}\vev{15}\vev{16}\vev{23}^2\vev{24}\vev{25}\vev{26}\vev{45}
\vev{46}\vev{56}}.}

Adding all four contributions, we get
\eqn\mhvsix{A(1^-,2^-,3^+,4^+,5^+,6^+)=D_1+D_1(4\leftrightarrow5)
+D_1(4\leftrightarrow6)+ D_2.} \mhvsix\ agrees with the known
result for the six graviton MHV amplitude.

The non-MHV amplitude $A(1^-,2^-,3^-,4^+,5^+,6^+)$ has
contribution from six classes of diagrams $D_1=
(2,\hat3|\hat4,5,6,1)+(1\leftrightarrow2),
D_2=(1,6,\hat3|\hat4,2,5)+(1\leftrightarrow2) +(5\leftrightarrow6)
+(1\leftrightarrow2,5\leftrightarrow6),
D_3=(2,5,6,\hat3|\hat4,1)+(1\leftrightarrow2), D_4=\bar
D_3^{flip}, D_5=\bar D_1^{flip}$ and $D_6=(5,6,\hat3|\hat4,1,2).$
The 'conjugate flip' $\bar D^{flip}$ exchanges the spinor products
$\vev{}\leftrightarrow []$ and the labels $i\leftrightarrow7-i.$

The first class of diagrams $D_1:
(2,\hat3|\hat4,5,6,1)+(1\leftrightarrow2)$ evaluates to
\eqn\dones{\eqalign{D_1=&{\vev{23}\langle1|2+3|4]^7\big(\langle1|2+3|4]
\langle5|3+4|2][51]+[12][45]\vev{51}p_{234}^2\big)\over
\vev{15}\vev{16}[23][34]^2\vev{56}p_{234}^2\langle1|3+4|2]\langle5|3+4|2]
\langle5|2+3|4]\langle6|3+4|2]\langle6|2+3|4]}\cr
&+(1\leftrightarrow2).}} The second group,
$D_2:(1,6,\hat3|\hat4,2,5)\, +$ permutations, gives
\eqn\dtwos{\eqalign{D_2=&-{\vev{13}^7\vev{25}[45]^7[16]\over
\vev{16}[24][25]\vev{36}p_{245}^2\langle1|2+5|4]\langle6|2+5|4]
\langle3|1+6|5]\langle3|1+6|2]}\cr &+(1\leftrightarrow2)
+(5\leftrightarrow6) +(1\leftrightarrow2,5\leftrightarrow6).}} The
third class $D_3:(2,5,6,\hat3|\hat4,1)+(1\leftrightarrow2)$ is
\eqn\dthrees{\eqalign{D_3=&{\vev{13}^8[14][56]^7\big(
\vev{23}\vev{56}[62]\langle1|3+4|5]+\vev{35}[56]\vev{62}
\langle1|3+4|2]\big)\over
\vev{14}[25][26]\vev{34}^2p_{134}^2\langle1|3+4|2]\langle1|3+4|5]
\langle1|3+4|6]\langle3|1+4|2]\langle3|1+4|5]\langle3|1+4|6]}\cr
&+(1\leftrightarrow2).}} The fourth and fifth group are related by
conjugate flip to the third and first group respectively. The last
group to evaluate consists of a single diagram
$D_6:(5,6,\hat3|\hat4,1,2)$ \eqn\dfives{D_6={\vev{12}[56]
\langle3|1+2|4]^8\over[21][14][24]\vev{35}\vev{36}\vev{56}p_{124}^2
\langle5|1+2|4]\langle6|1+2|4]\langle3|5+6|1]\langle3|5+6|2]}.}
Adding the pieces together, the six graviton non-MHV amplitude
reads \eqn\sixnon{A(1^-,2^-,3^-,4^+,5^+,6^+)=D_1+\bar D_1^{flip}+
D_2+D_3+\bar D_3^{flip}+D_6.}


\newsec{Derivation of the Recursion Relations}

The derivation of the tree-level recursion relations \grtree\
goes, with few modifications, along the same lines as the
derivation of the tree-level recursion relations for scattering
amplitudes of gluons \BrittoFQ,\ so we will be brief.

We start with the scattering amplitude $A(z)$ defined at shifted
momenta, see \zfunction\ and \shiftp.\ $A(z)$ is a rational
function of $z$ because the $z$ dependence enters the scattering
amplitude only via the shifts $\tilde\lambda_i\rightarrow
\tilde\lambda_i+z\tilde\lambda_j$ and $\lambda_j\rightarrow
\lambda_j-z\lambda_i$ and because the original tree-level
scattering amplitude is a rational function of the spinors.

Actually, for generic momenta, $A(z)$ has only single poles in
$z.$ These come from the singularities of the propagators in
Feynman diagrams. To see this, recall that for tree level
amplitudes, the momentum through a propagator is always a sum of
momenta of external particles $P_{\cal
I}=p_{i_1}+p_{i_2}+\dots+p_{i_l},$ where ${\cal I}$ is a group of
${\it not}$ necessarily adjacent gravitons. At nonzero $z,$ the
momentum becomes $P_{\cal
I}(z)=p_{i_1}(z)+p_{i_2}(z)+\dots+p_{i_l}(z).$ Here, $p_k(z)$ is
independent of $z$ for $k\neq i,j$ and $p_i(z)+p_j(z)$ is
independent of $z.$ Hence, $P_{\cal I}(z)$ is independent of $z$
if  both $i$ and $j$ are in ${\cal I}$ or if neither of them is in
${\cal I}.$ In the remaining case, one of $i$ and $j$ is in the
group ${\cal I}$ and the other is not. Without loss of generality,
we take $i\in {\cal I}.$  Then $P_{\cal I}(z)=P_{\cal I}
+z\lambda_i\tilde\lambda_j$ and $P_{\cal I}^2(z)=P_{\cal
I}^2-z\langle i|P_{\cal I}|j].$ Clearly, the propagator $1/P_{\cal
I}(z)^2$ has a simple pole for \eqn\zi{z_{\cal I}={P_{\cal
I}^2\over\langle i|P_{\cal I}|j]}.} For generic momenta,  $P_{\cal
I}$'s are distinct for distinct groups ${\cal I},$ hence the
$z_{\cal I}$'s are distinct. So all singularities of $A(z)$ are
simple poles.

To continue the argument, we need to assume that $A(z)$ vanishes
as $z\rightarrow \infty.$ In the next section we will argue that
the tree level graviton amplitudes obey this criterium.  A
rational function $A(z)$ that has only simple poles and vanishes
at infinity can be expressed as \eqn\rata{A(z)=\sum_{\cal I}{{\rm
Res}\, A(z_{\cal I})\over z-z_{\cal I}},} where ${\rm Res}
A(z_{\cal I})$ are the residues of $A(z)$ at the simple poles
$z_{\cal I}.$ The physical scattering amplitude is simply $A(0)$
\eqn\physa{A=-\sum_{\cal I} {{\rm Res}\, A(z_{\cal I})\over
z_{\cal I}}.} It follows from  the above discussion that the sum
is over ${\cal I}$ such that $i$ is in ${\cal I}$ while $j$ is
not.

The residue ${\rm Res}\, A(z_{\cal I})$ has contribution from
Feynman diagrams which contain the propagator $1/P_{\cal I}^2.$
The propagator divides the tree diagram into ``left" part
containing gravitons in ${\cal I}$ and ``right" part containing
gravitons in ${\cal J}=(1,2,\dots,n)-{\cal I}.$ For $z\rightarrow
z_{\cal I},$ the propagator with momentum $P_{\cal I}$ goes
on-shell and the left and right part of the diagram approach
tree-level diagrams for on-shell amplitudes. The contribution of
these diagrams to the pole is \eqn\pole{ \sum_h A^h_L(z_{\cal I})
{1\over P_{\cal I}^2(z)} A^{-h}_R(z_{\cal I}),} where the sum is
over the helicity $h=\pm$ of the intermediate graviton.  This
gives the recursion relation \grtree.\


\newsec{Large $z$ Behavior of Gravity Amplitudes}

To complete the proof, it remains to show that the amplitude
$A(z)$ goes to zero as $z$ approaches infinity. We were able to
obtain only partial results in this direction, which we now
discuss.

\subsec{Vanishing of the MHV Amplitudes}

Let us firstly consider the large $z$ behavior of the $n$ graviton
MHV amplitude \BerendsZP\
\eqn\grmhv{\eqalign{A(1^-,2^-,3^+,\dots,n)&={\vev{12}^8 }\Big\{
{[23]\langle n|P_{2,3}|4]\langle n|P_{2,4}|5] \dots \langle n
|P_{2,n-2}|n-1] \over \vev{12}\vev{23}\dots \vev{n-2,n-1}
\vev{n-1,1}\vev{1n}^2\vev{2 n}\vev{3 n}\dots \vev{n-1,n}} \cr
&\quad \quad \qquad +{\rm \, permutations \,\, of \,\,} (3,4,\dots
, n-1) \Big\},}} where $P_{i,j}=\sum_{k=i}^j p_k.$ The formula is
valid for $n\geq 5.$ It follows from supersymmetric Ward
identities that the expression in the bracket is totally
symmetric, although this is not manifest.

The terms in the curly brackets are completely symmetric  so they
contribute the same power of $z$ independently of $i$ and $j.$ To
find the contribution of the terms in the brackets, we pick a
convenient value $(i,j)=(1,n)$ for the reference gravitons. Recall
that $\tilde\lambda_i(z)$ and $\lambda_j(z)$ are linear in $z$
while $\lambda_i$ and $\tilde\lambda_j$ do not depend on $z.$ It
follows that the numerator of each term in the brackets \grmhv\
goes like $z^{n-4}$ and the denominator gives a factor of
$z^{n-2}.$ Hence, the terms in the brackets give a factor of
$1/z^2.$ This factor is the same for all choices of reference
momenta by the complete symmetry of the terms in the brackets. For
the helicity configurations $(h_i,h_j)=(-,+),(+,+)$ and $(-,-),$
the factor $\vev{12}^8$ does not contribute, so the amplitude
vanishes at infinity as $A_{MHV}\sim {1/ z^2}.$

A recent paper \NairGR\ relates the MHV amplitudes to current
correlators on curves in twistor space. This raises the
possibility of a twistor string description of perturbative ${\cal
N}=8$ supergravity \WittenNN.\  In the gauge theory case, the
twistor string leads to an MHV diagrams construction for the tree
level scattering amplitudes \CachazoKJ.\ One computes the
tree-level amplitudes from tree-level Feynman diagrams in which
the vertices are MHV amplitudes, continued off-shell in a suitable
manner, and the propagators are ordinary Feynman propagators.

The vanishing of the gluon scattering amplitude $A(z)$ at infinity
follows very easily from the vanishing of the MHV diagrams via the
MHV diagrams construction \BrittoFQ.\ We would like to speculate,
that it might be possible to prove the vanishing of graviton
scattering amplitude $A(z)$ along the same lines using the
hypothetical MHV diagrams construction.

\subsec{Analysis of the Feynman Diagrams}

In this section we study the large $z$ behavior of Feynman
diagrams contributing to $A(z)$ following \BrittoFQ.\

Recall that any Feynman diagram contributing to $A(z)$ is linear
in the polarization tensors $\epsilon_{a\dot a, b\dot b}$ of the
external gravitons. The polarization tensors of all but the
$i^{th}$ and $j^{th}$ graviton are independent of $z.$ To find the
$z$ dependence of the polarization tensors of the reference
gravitons, recall that $\tilde\lambda_i(z),\lambda_j(z)$ are
linear in $z$ and $\lambda_i,\tilde\lambda_j$ do not depend on
$z.$ It follows from \grpol\ and \glpol\ that the polarization
tensors of the reference gravitons give a factor of $z^{\pm 2}$
depending on their helicities. Hence, the polarization tensors can
suppress $A(z)$ by at most a factor of $z^4.$

The remaining pieces in Feynman diagrams are constructed from
vertices and propagators that connect them. Perturbative gravity
has infinite number of vertices coming from the expansion  of the
Einstein-Hilbert Lagrangian \eqn\einstein{{\cal L}=-\sqrt{-g}\,R}
around the flat vacuum $g_{\mu\nu}=\eta_{\mu\nu}+h_{\mu\nu}.$ The
graviton vertices have two powers of momenta coming from the two
derivatives in the Ricci scalar.

The $z$ dependence in a tree level diagram "flows" along a unique
path of Feynman propagators from the $i^{th}$ to the $j^{th}$
graviton. In a path composed of $k$ propagators, there are $k+1$
vertices. Each propagator contributes a factor of $1/z$ and each
vertex contributes a factor of $z^2.$ Altogether, the propagators
and vertices give a factor of $z^{k+2}.$

The product of polarization tensors vanishes at best as $1/z^4,$
so the contribution of individual Feynman diagrams to $A(z)$ seems
to grow at infinity as $z^{k-2},$ where $k$ is the number of
propagators from the $i^{th}$ to the $j^{th}$ graviton. Clearly,
in a generic Feynman diagram, this number grows with the number of
external gravitons. So this analysis suggests that $A(z)$ grows at
infinity with a power of $z$ that grows as we increase the number
of external gravitons.

This is in contrast to the above analysis of MHV amplitude  that
vanishes at infinity as $1/z^2.$ The vanishing of $A(z)$ at
infinity depends on unexpected cancellation between Feynman
diagrams.

\subsec{KLT Relations and the Vanishing of Gluon Amplitudes}

A different line of attack is to express the graviton scattering
amplitudes via the KLT relations in terms of the gluon scattering
amplitudes. One then infers behavior of $A(z)$ at infinity from
the known behavior \BrittoFQ\ of the gauge theory amplitudes. The
KLT relations have been used in past to show that ${\cal N}=8$
supergravity amplitudes \refs{\CremmerKM,\CremmerDS} have better
than expected \refs{\BernUG, \HoweUI} ultraviolet behavior so we
expect that the KLT relations give us a better bound on $A(z)$
than the analysis of Feynman diagrams. Indeed, we will use them to
prove the vanishing of $A(z)$ at infinity up to six gravitons and
up to eight gravitons in the appendix.

The KLT relations of string theory \KawaiXQ\ relate the closed
string amplitudes to the open string amplitudes. They arise from
representing each closed string vertex operator $C$ as the product
of two open string vertex operators $O$
\eqn\openclosed{C(z_i,{\bar z}_i)=O(z_i){\bar O}({\bar z}_i)} and
deforming the closed string integration contour into two sets of
open string integration contours.

In the infinite tension limit, the KLT relations relate the
gravity amplitudes to the gauge theory amplitudes \BerendsZP.\ The
tree level  KLT relations up to six gravitons are
\eqn\klt{\eqalign{A(1,2,3)&=\CA(1,2,3)^2\cr
A(1,2,3,4)&=s_{12}\CA(1,2,3,4)\CA(1,2,4,3)\cr
A(1,2,3,4,5)&=s_{12}s_{34}\CA(1,2,3,4,5)
\CA(2,1,4,3,5)+s_{13}s_{24} \CA(1,3,2,4,5)\CA(3,1,4,2,5)\cr
A(1,2,3,4,5,6)&= s_{12} s_{45} \CA(1,2,3,4,5,6) \big\{ s_{35}
\CA(2,1,5,3,4,6)+(s_{34}+s_{35})\CA(2,1,5,4,3,6)\big\}\cr & \quad
+ {\rm permutations\,\, of\,\, } (234),}} where
$s_{ij}=(p_i+p_j)^2.$ $A(1,2,\dots, n)$ is the $n$ graviton
scattering amplitude and $\CA(i_1,i_2,\dots,i_n)$ is the color
ordered gauge theory amplitude. Each graviton state on the left is
the product of two gauge theory states on the right. The
decomposition of the graviton states comes from the infinite
tension limit of the decomposition of the closed string vertex
operators \openclosed.\ It is reflected in the factorization of
the graviton polarization tensor \grpol\ in terms of the gluon
polarization vectors. The KLT relations for any number of
gravitons are written down in Appendix A of \BernSV\ and
schematically in the appendix.

The KLT relations express an $n$ graviton scattering amplitude as
a sum of products of two gluon scattering amplitudes and $n-3$
$s_{ij}$ invariants. The gluon scattering amplitudes vanish at
infinity as $1/z$ or faster \BrittoFQ.\ Hence, KLT relations imply
the vanishing at infinity of the graviton amplitudes as long as
the products of $s_{ij}$'s in \klt\ grow at most linearly with
$z.$

For $n\leq 6$ gravitons, a quick glance at \klt\ shows that this
is the case. We rename the gravitons so that the reference
gravitons are $1$ and $n.$ The products of $s_{ij}$'s in \klt\ are
independent of $p_n$ and linear in $p_1.$ Hence they give one
power of $z$ because $p_1(z)$ and $p_n(z)$ are linear in $z$ and
$p_k$ for $k\neq 1,n$ is independent of $z.$ It follows that
$A(z)$ vanishes as $1/z$ or faster as $z\rightarrow \infty$ for
less than seven gravitons.

For seven or more gravitons, an analysis of the general KLT
relations shows that on the right hand side of KLT relations,
there are always some products of $n-3$ $s_{ij}$'s that have more
than one power of the reference momenta. The corresponding terms
in the KLT relations are not expected to vanish at infinity.
Hence, the function $A(z)$ does not vanish at infinity unless
there is an unexpected cancellation between different terms in the
KLT relations.

In the appendix we present a more careful study of KLT relations
that reveals that $A(z)$ vanishes for $n\leq 8$.

\subsec{Proof of Vanishing of $A(z)$ for NMHV Amplitudes}

NMHV amplitudes are those with three negative helicity gravitons
and any number of plus helicity gravitons,
$A(p_1^-,p_2^-,p_3^-,p_4^+,\ldots, p_n^+)$.

Consider the following function of $z$,
$A_a(p_1^-(z),p_2^-,p_3^-,p_4^+(z),\ldots, p_n^+(z))$, where
\eqn\aux{ p_1(z) = \lambda_1 \left( \tilde\lambda_1 + z
\sum_{i=4}^n \tilde\lambda_i \right), \qquad p_k(z) = (\lambda_k -
z\lambda_1)\tilde\lambda_k }
for $k=4,\ldots ,n$. The subscript $a$ in $A_a(z)$ stands for
auxiliary. The idea is to derive an new set of recursion relations
for $A_a(z)$ which we use later on to prove that $A(z)$ vanishes
at infinity.

In order to get the auxiliary recursion relations we start by
proving from Feynman diagrams that $A_a(z)$ vanishes as $z\to
\infty.$  Note that $(n-2)$ polarization tensors depend on $z$ and
with the choice made in \aux\ all of them vanish as $1/z^2$. The
most dangerous Feynman diagram is the one with the largest number
of vertices. Such a diagram must only have cubic vertices. For $n$
gravitons there are $n-2$ vertices. Each vertex contributes a
factor of $z^2$. Altogether, the polarization tensors contribute a
factor of $1/z^{2(n-2)}$ and the vertices contribute a factor of
$z^{2(n-2)}$ which gives a constant for large $z$. Now we have to
consider propagators. Each propagator that depends on $z$ goes
like $1/z$. Therefore, all we need is that in every diagram at
least one propagator depends on $z$. From \aux\ it is easy to see
that the only propagator that does not depend on $z$ is
$1/(p_2+p_3)^2$. A diagram with only this propagator has exactly
two vertices and therefore our proof is complete for $n>4$.

The shift in \aux\ can be thought of as iterating  the shift
introduced in \BrittoFQ.\foot{This iteration procedure was used
recently in \BernHS\ to find recursion relations for all plus
one-loop amplitudes of gluons.}
 Now we can follow the same steps
as in section 4 to derive recursion relations based on the pole
structure of $A_a(z)$.

We find
\eqn\nola{ A_a(z) = \sum_{\cal I}\sum_h A_{\cal I}(z_{\cal
I},P^h_{\cal I}(z_{\cal I})){1\over P^2_{\cal I}(z)} A_{\cal
J}(z_{\cal I},-P^{-h}_{\cal I}(z_{\cal I})).}
where the sum is over all possible sets of two or more gravitons
$\CI\neq\{2,3\},$ such that the graviton $1$ is not in $\CI.$
Here, $\CJ$ is the complement of $\CI.$

The main advantage of choosing the same negative helicity graviton
in \aux\ to pair up with all plus helicity gravitons is that
$P^2_{\cal I}(z)$ is a linear function of $z$. Therefore, the
location of the poles $z_{\cal I}$ is easily computed to be of the
form
\eqn\holu{ z_{\cal I} = {P_{\cal I}^2 \over \sum_j \langle
1|P_{\cal I} | j] } }
where the sum in $j$ runs over all gravitons in ${\cal I}$ that
depend on $z$.

Setting $z$ to zero in \nola\ gives us a new representation of the
original amplitude, i.e.,
\eqn\newr{A(p_1^-,p_2^-,p_3^-,p_4^+,\ldots, p_n^+) = \sum_{\cal
I}\sum_h A_L(z_{\cal I},P^h_{\cal I}(z_{\cal I})){1\over P^2_{\cal
I}} A_R(z_{\cal I},-P^{-h}_{\cal I}(z_{\cal I})).  }

This is a new set of recursion relations for NMHV amplitudes.
However, the expressions obtained from \newr\ are naturally more
complicated than the ones obtained from the one introduced in
section 2. Instead of computing amplitudes with \newr, the idea is
to use it to prove that $A(z)$ of section 2 vanishes for large
$z$.

Consider $A(z)$ constructed from \newr\ by defining
\eqn\olds{ p_1(z) = \lambda_1(\tilde\lambda_1 + z
\tilde\lambda_4), \qquad p_4(z) = (\lambda_4 - z
\lambda_1)\tilde\lambda_4.}

There are two different kind of terms in \newr. One class consists
of those where $p_1$ and $p_4$ are on the same side. This implies
that neither $P_{\cal I}$ nor $z_{\cal I}$ depends on $z$.
Therefore, the $z$ dependence is confined into one of the
amplitudes, say $A_L$. But this is an amplitude with less
gravitons and by induction we assume that it vanishes for large
$z$.

The second class of terms is more subtle. Since $p_1(z)$ and
$p_4(z)$ are on different sides, both $P_{\cal I}$ and $z_{\cal
I}$ become functions of $z$.

It turns out that $z_{\cal I}$ is a linear function of $z$. More
explicitly\foot{Had we chosen a different negative helicity
graviton in \olds, we would have found that $z_{\cal I}$ becomes a
rational function of $z$.},
\eqn\more{ z_{\cal I}(z) = {P_{\cal I}^2 + z \langle 1| P_{\cal
I}| 4 ] \over \sum_j \langle 1|P_{\cal I} | j] }. }
Recall that $P_{\cal I}$ denotes $P_{\cal I}(0)$.

Now we are left with $A_{\cal I}$ and $A_{\cal J}$ in \newr, one
with $n_1+1$ and the other with $n_2+1$ gravitons. Note that
$n=n_1+n_2$. In a Feynman diagram expansion of each of them we can
single out the most dangerous diagrams and multiply them to get
the most dangerous terms in \newr. Each diagram contributes a
factor of $z^{2(n_i-1)}$ from the vertices. Therefore we find
$z^{2(n-2)}$. From the polarization tensors we find $z^{-2(n-2)}$,
this comes from the $z$ dependence of $z_{\cal I}$. The
polarization tensors for the internal gluons give a factor of
\eqn\pof{\sum_h \epsilon^h_{\mu\nu} \epsilon^{-h}_{\lambda\rho}=
d_{\mu\rho}d_{\nu\lambda}+d_{\mu\lambda}d_{\nu\rho}-d_{\mu\nu}d_{\rho\lambda},}
where \eqn\dfa{d_{\mu\nu}=\eta_{\mu\nu}-{k_\mu n_\nu+k_\nu
n_\mu\over k\cdot n}.} Here, $k= P_{\cal I}(z)$ is the momentum of
the internal propagator and $n_{a\dot a} =\mu_a\tilde\mu_{\dot a}
$ is an auxiliary vector used in the definition \glpol. $n_{a\dot
a}$ is taken no collinear with $k$. For large $z,$ the tensor
$d_{\mu\nu}$ does not depend on $z$ so the polarization tensors of
the internal gravitons do not contribute a factor of $z.$

Finally, the propagator in \newr\ is $1/P^2_{\cal I}(z)$, which
vanishes as $1/z$. Therefore, the most dangerous term in $A(z)$
vanishes as $1/z$.

For large $z$ the $z$-dependence of $P_{\cal I}(z_{\cal I})$ in
\newr\ can be made trivial, for the product $A_{\cal I}A_{\cal J}$ must be homogeneous
of degree zero under scalings of $P_{\cal I}(z_{\cal I})$.

This completes the proof of the recursion relations of section 2
for next-to-MHV amplitudes of gravitons.

While we are not able to prove that $A(z)$ vanishes at infinity
for general amplitudes with more than eight gravitons, we showed
above that $A(z)$ vanishes at infinity for MHV and NMHV amplitudes
with arbitrary number of gravitons. Hence, the recursion relations
are valid for all MHV and NMHV amplitudes contrary to the
expectations from KLT relations. This raises the hope, that the
recursion relations are valid for other scattering amplitudes of
gravitons as well. In particular, one might expect that by
considering more general auxiliary recursion relations one could
prove that $A(z)$ vanishes at infinity for general gravity
amplitudes.

\vskip 0.5cm

{\bf Note:}

After this work was completed, \BedfordYY\ appeared which has some
overlap with our results.

\vskip 1cm

\centerline{\bf{ Acknowledgements}}

It is pleasure to thank R. Britto, B. Feng, K. Intriligator and E.
Witten for useful discussions. Work of F. Cachazo was supported in
part by the Martin A. and Helen Chooljian Membership at the
Institute for Advanced Study and by DOE grant DE-FG02-90ER40542
and that of P. Svr\v{c}ek in part by Princeton University
Centennial Fellowship and by NSF grants PHY-9802484 and
PHY-0243680. Opinions and conclusions expressed here are those of
the authors and do not necessarily reflect the views of funding
agencies.


\appendix{A}{Proof of Vanishing of $A(z)$ for $z\rightarrow \infty$ up to Eight Gravitons.}

In this appendix we provide further evidence for validity of the
recursion relations \grtree.\ We show that the recursion relations
hold for any graviton amplitude up to eight gravitons. Recall that
we need to prove that the auxiliary function $A(z)$ \zfunction\
vanishes at infinity. We demonstrate this using the KLT relations.

The basic fact we will use is that the function $\CA(z)$ for a
gluon scattering amplitude goes like $1/z^2$ at infinity for
non-adjacent marked gluons. Hence, picking the marked gluons so
that they are non-adjacent in all terms in KLT relations, the
product of two Yang-Mills amplitudes in each term goes like
$1/z^4.$

Hence, $A(z)$ vanishes at infinity as long as the products of
$s_{ij}$'s in the KLT relations do not contribute more than a
factor of $z^3.$ An inspection of the KLT relations will show that
this holds at least up to eight gravitons which will complete the
proof.

Let us begin by showing that the gluon amplitudes go as $1/z^2$ as
$z\rightarrow\infty$ for non-adjacent marked gluons with
helicities $(h_i,h_j)=(+,+),(-,-),(-,+).$ The argument uses MHV
rules and is a simple generalization of the argument given in
\BrittoFQ,\ which showed that the amplitudes vanish as $1/z.$ We
assume that $h_j=+.$ For $h_j=-$ one makes the same argument using
the opposite helicity MHV rules.

Firstly, consider the $n$ gluon MHV amplitude
\eqn\mhvampl{\CA(r^-,s^-)={\vev{r,s}^4\over \prod_{k=1}^n
\vev{k,k+1}}.} Recall that $\lambda_j(z)=\lambda_j-z\lambda_i$ is
linear in $z$ and $\lambda_i(z)=\lambda_i$ is independent of $z.$
For $h_j=+,$ $\lambda_j$ does not occur in the numerator. In the
denominator it appears in the two factors
$\vev{\lambda_{j-1},\lambda_j}$ and
$\vev{\lambda_j,\lambda_{j+1}},$ both of which are linear in $z$
for $i$ not adjacent to $j.$ Hence for $h_i=+$ and $|i-j|>1,$ the
MHV amplitude goes like $1/z^2$ at infinity.

For general amplitudes, we use MHV diagram constructions. In this
construction, the amplitudes are built from Feynman vertices which
are suitable off-shell continuations of the MHV amplitudes. The
vertices are connected with ordinary scalar propagators.

The Feynman vertices are the MHV amplitudes \mhvampl,\ where we
take $\lambda^a=P^{a\dot a}\eta_{\dot a}$ for an off-shell
momentum $P.$ Here $\eta$ is an arbitrary positive helicity
spinor. The physical amplitude, which is a sum of MHV diagrams, is
independent of the choice of $\eta$ \CachazoKJ.\

The internal momentum $P$ can depend on $z$ only through a shift
by the null vector $z\lambda_i\tilde\lambda_j.$ Taking
$\eta=\tilde\lambda_j,$  $\lambda^a=P^{a\dot a}\tilde\lambda_{j\,
\dot a}$ becomes independent of $z.$ Hence, the internal lines do
not introduce additional $z$ dependence into the MHV vertices. The
MHV vertices give altogether a factor of $1/z^2$ from the two
powers of $\lambda_j(z)$ in the denominator of one of the
vertices. The propagators $1/k^2$ are either independent of $z$ or
contribute a factor of $1/z.$ So, a general gluon amplitude goes
like $1/z^2$ at infinity for non-adjacent marked gluons.

The KLT relations \BernSV\ for $n$ gravitons are
\eqn\kltn{\eqalign{A(1,2,\dots,n)=&\Big(\CA(1,2,\dots,n)
\sum_{perm}f(1,i_1,\dots,i_j)\bar f(n-1,l_1,\dots, l_{j'}) \cr
&\times \CA(i_1,\dots, i_j,1, n-1,l_1,\dots,l_{j'},n)\Big) \cr &+
\CP(2,\dots,n-2),}} where $j=\lfloor n/2\rfloor-1, j'=\lfloor
(n-1)/2\rfloor-1$ and the permutations are $(i_1,\dots, i_j)\in
\CP(2,\dots, \lfloor n/2\rfloor )$ and $(l_1,\dots,l_{j'})\in
\CP(\lfloor n/2 \rfloor+1,\dots,n-2).$ The exact form of the
functions $f$ and $\bar f$ does not concern us here. The only
property we need is that $f$ and $\bar f$ are homogeneous
polynomials of degree $j$ and $j'$ in the Lorentz invariants
$p_m\cdot p_n$ with $m,n\in(1,i_1,\dots,i_j)$ or $m,n\in
(l_1,\dots,l_{j'},n-1)$ respectively.

Consider $A(z)$ with marked gravitons $n$ and $k$ where $k$ is any
label from the set $(2,\dots,n-2).$ In the KLT relations \kltn,\
the gluon amplitude $\CA(1,2,\dots, n)$ contributes a factor of
$1/z^2$ since $k$ and $n$ are non-adjacent. For $k\in (i_1,\dots,
i_j)$ the second gluon amplitude gives a factor of $1/z^2$ and $f$
gives at most a factor of $z^j,j=\lfloor n/2\rfloor-1.$ Hence, the
terms with $k\in (i_1,\dots, i_j)$ are bounded at infinity by
$z^{\alpha}$ where $\alpha=\lfloor n/2\rfloor -5.$ For $k\in
(l_1,\dots, l_{j'})$ the second gluon amplitude gives a factor of
$1/z$ because $k$ and $n$ might be adjacent. $\bar f$ contributes
$z^{j'}, j'=\lfloor(n-1)/2\rfloor-1$ so the graviton amplitude is
bounded by $z^{\alpha'}, \alpha'=\lfloor (n-1)/2\rfloor-4.$ The
exponents $\alpha,\alpha'$ are negative for $n\leq 8,$ which
completes the proof of the recursion relations up to eight
gravitons.

\listrefs
\end